\documentclass[twocolumn]{jpsj2}
\usepackage{graphics}

\author{%
Norikazu \textsc{TODOROKI}\thanks{E-mail
address: todoroki@kanagawa-u.ac.jp} }

\title{Sublattice Asymmetric Reductions of Spin Values on Stacked Triangular Lattice Antiferromagnet CsCoBr$_3$}

\inst{Institute of Physics, Kanagawa University,
3-27-1, Rokkaku-bashi, 
Kanagawa-ku, Yokohama, Kanagawa 221-8686 Japan.
,
}

\recdate{\today}

\abst{%
We study the reductions of spin values of the ground state on a stacked triangular antiferromagnet using the spin-wave approach. 
We find that the spin reductions have sublattice 
asymmetry due to the cancellation of the molecular field.
The sublattice asymmetry qualitatively analyzes the NMR results 
of CsCoBr$_3$. 
}

\kword{%
CsCoBr$_3$, NMR, quantum fluctuation, frustration
}

\begin{document}
\sloppy
\maketitle
\section{Introduction} 
Hexagonal ABX$_3$ type crystals, such as CsCoBr$_3$ and CsCoCl$_3$, are an interesting subject in studies on rare magnetic ordered phases and successive phase transitions. 
The Co$^{2+}$ magnetic ions have fictitious $S=1/2$ Ising-like spin
and form a stacked triangular lattice.
These substances have a strong geometrical frustration since the spins are antiferromagnetically coupled in the c-plane.
Further, these substances have been studied as 
quasi-one-dimensional substances for a long time
because the chains along the $c$-axis are mutually
 separated from each other by large Cs$^{+}$ ions; 
this results in an interchain exchange that is about 2 or 3 orders of 
magnitude smaller than the intrachain interaction.\cite{mekata}

\begin{figure}
\begin{center}
\includegraphics[scale=0.7]{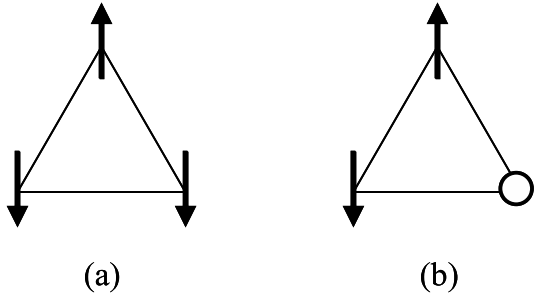}
\caption[]{Schematic sublattice magnetizations of (a) FR phase and (b) PD phase. The arrows and circle denote the sublattice magnetizations. \label{fig1}}
\end{center}
\end{figure}

The magnetic ordering of these substances is explained 
by a stacked triangular Ising model with weak next nearest neighbor interaction
in the $c$-plane.
This model has two successive phase transitions
and two types of ordered phases.\cite{todoroki}
As the temperature is reduced, there appear a partial disordered (PD) phase and a ferrimagnetic (FR) phase which are schematically depicted in Fig.~\ref{fig1}.
In the PD phase, the spins on the two sublattices are antiferromagnetically ordered, while those on the other sublattice are disordered.
In the FR phase, the spins on the two sublattices are ordered in the same direction, while those 
and the spins on the other sublattice are ordered in the opposite direction.

\begin{figure}
\begin{center}
\includegraphics[scale=0.7]{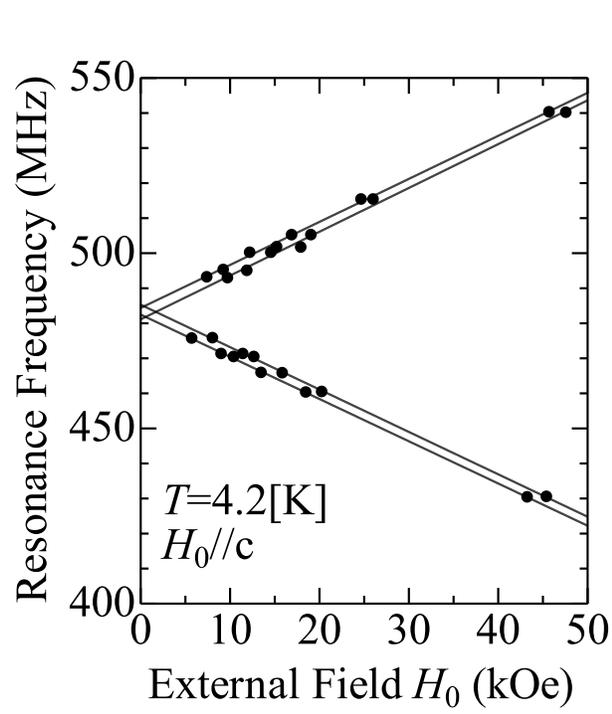}
\caption[]{External field $H_0$ dependence of Co NMR frequencies which is taken from Ref. \citen{uyeda}. \label{fig2}}
\end{center}
\end{figure}

\begin{figure}
\begin{center}
\includegraphics[scale=0.7]{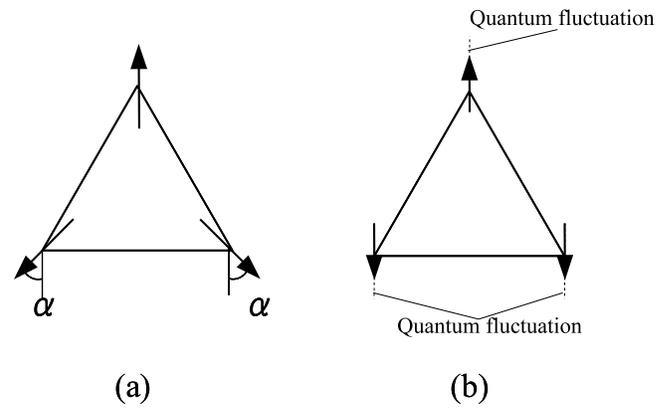}
\caption[]{(a) Canted three-sublattice model. (b) Present spin reduction model. \label{fig3}}
\end{center}
\end{figure}

Recently, Uyeda et al. investigated the FR ground state of CsCoBr$_3$ by a $^{59}$Co nuclear spin ($S=7/2$) echo spectra at $T=4.2$ [K] using a single crystal. \cite{uyeda}
They explained that the obtained NMR spectrum is the composition of two 
spectra with an intensity ratio 1:2.
Fig.~\ref{fig2} shows the external field $H_0$ dependence of Co NMR frequencies which is taken from Ref. \citen{uyeda}. 
The resonance frequencies at $H_0=0$ indicates $\nu_N^1=484.7\pm 0.5$ [MHz] and $\nu_N^2=481.6\pm 0.5$ [MHz].
As a result, The hyperfine fields are estimated as $H_N^1=480.0\pm 0.5$ [kOe] and $H_N^2=477.0\pm 0.5$ [kOe] 
by using the free $^{59}$Co's gyromagnetic ratio $\gamma=1.01$ [MHz/kOe].
This behavior is significantly different from that revealed in the NMR results of CsCoCl$_3$.
With regard to CsCoCl$_3$, the NMR spectrum shows eight sharp peaks.
The results of $^{59}$Co NMR in CsCoBr$_3$ are summarized in Table \ref{table1} along with that of CsCoCl$_3$ from Refs. \citen{uyeda} and \citen{tkubo}.
In order to explain the NMR results, Uyeda et al. proposed a canted three-sublattice model
where the spins on the two parallel sublattices are canted and
the spins on the single opposite sublattice do not cant.
The schematic figure of the canted three-sublattice model is shown in Fig.~\ref{fig3} (a).
From the ratio of $H_{N}^1$ and $H_N^{2}$, 
the cant angle of the two parallel sublattice moment is estimated  as $\alpha\sim 6$ [deg] in the ground state of CsCoBr$_3$.
Moreover, Uyeda et al. investigated the $^{79,81}$Br nuclear spin echo spectra and they obtained the result that could explain by the canted three-sublattice model. \cite{uyeda2}
Although the NMR spectra can be explained by the canted three-sublattice model, the mechanism of canting has not been understood yet.

In the present paper, we consider the quantum fluctuation
along the $c$-axis and calculate the spin reductions on the ground state (Fig.~\ref{fig3} (b)).
The sublattice asymmetry of the spin reductions is observed.
We expect that this sublattice asymmetry explains the NMR results of CsCoBr$_3$.

\section{Model}
The spin Hamiltonian of these substances are expressed to be,
\begin{eqnarray}
&&{\cal H}=-2J_0\sum_{i}\left \{ S_{i,\mu}^zS_{i+1,\mu}^z
+\epsilon \left (S_{i,\mu}^xS_{i+1,\mu}^x+S_{i,\mu}^yS_{i+1,\mu}^y \right )
\right \} \nonumber \\
&&-2J_1\sum_{\langle \mu\nu\rangle}^{nn}\left \{ S_{i,\mu}^zS_{i,\nu}^z
+\epsilon \left (S_{i,\mu}^xS_{i,\nu}^x+S_{i,\mu}^yS_{i,\nu}^y \right )
\right \} \nonumber \\
&&-2J_2\sum_{\langle \mu\nu\rangle}^{nnn}\left \{ S_{i,\mu}^zS_{i,\nu}^z
+\epsilon \left (S_{i,\mu}^xS_{i,\nu}^x+S_{i,\mu}^yS_{i,\nu}^y \right )
\right \},
\end{eqnarray}
where the first summation is taken over the nearest-neighbor pairs along the $c$-axis, 
the second and third summations
are taken over the nearest-neighbor and second-neighbor pairs in the c-plane
of the stacked triangular lattice, respectively.
The nearest-neighbor interactions along the $c$-axis and in the c-plane
are antiferromagnetic. The second-neighbor interaction 
is ferromagnetic. For CsCoBr$_3$, 
the values of the exchange interactions are estimated as 
$J_1\sim -80$ [K], $J_1/J_0= 1.0\times 10^{-2}\sim 6.0\times 10^{-2}$, $J_2/J_0\sim -1.0\times 10^{-4}$ and 
$\epsilon=0.1\sim 0.2$ based on experiments. \cite{interaction,epsilon}

The quantum fluctuation along the $c$-axis is weaker than that in the $c$-plane.
Therefore, we can consider only the quantum fluctuation along the $c$-axis.
In order to observe the effect of the quantum fluctuation 
along the $c$-axis, we apply the
chain mean field approximation,
in other words, we focus on one chain and consider the other chains to 
be molecular fields. 
\begin{eqnarray}
 {\cal H}_A&=&-2J_0\sum_{i}\big \{ S_{i,\mu}^zS_{i+1,\mu}^z\nonumber \\
&+&\epsilon \left (S_{i,\mu}^xS_{i+1,\mu}^x+S_{i,\mu}^yS_{i+1,\mu}^y
\right )\big \}
-\sum_i (-1)^i H_{\rm eff}^{\rm A}S_i^z, \nonumber \\
\end{eqnarray}
where 
\begin{eqnarray}
H_{\rm eff}^{\rm A}=6J_1(m_{\rm B}+m_{\rm C})-12J_2m_{\rm A},
\end{eqnarray}
and subscripts A, B and C denote each sublattice. 
The order parameters are defined as follows:
\begin{eqnarray}
m_\lambda=\frac{3}{N}\left \langle\sum_{i\in \lambda} (-1)^iS_i\right \rangle, 
\hspace{1cm} (\lambda=\mbox{A, B and C}).
\end{eqnarray}
Here, we assume the FR ground state without the spin canting.
\section{Spin-Wave Approximation}

\begin{figure}
\begin{center}
\includegraphics[scale=0.5]{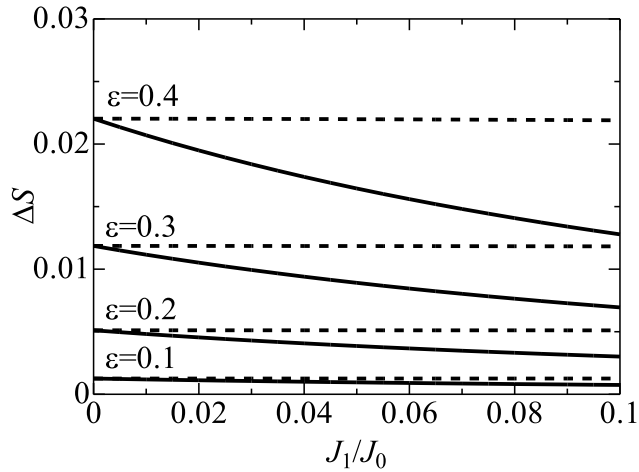}
\caption[]{$J_1/J_0$ and $\epsilon$ dependence of the spin reductions of each sublattice. The solid line represents the spin reductions of the single opposite sublattice, and the dashed line represents the spin reductions of the two parallel sublattices.\label{fig4}}
\end{center}
\end{figure}

The chain is divided into the $\alpha$ sublattice and $\beta$ sublattice. On the $\alpha$ ($\beta$) sublattice, the vacuum state is $S^{z}=S$ ($-S$).
In order to observe the spin reductions on the ground state by quantum fluctuation, 
we apply the spin-wave approach. \cite{kubo}
With Holstein-Primakoff transformation and Bogoliubov transformation,
we derive the diagonal Hamiltonian to be,
\begin{eqnarray}
&&{\cal H}=zJ_0NS^2-H_{\rm eff}^{\rm A}NS \nonumber \\
&&\hspace{-1cm}-2zJ_0S\sum_k\left \{ \sqrt{(1+h_{\rm eff}^{\rm A})^2-\epsilon^2\cos^2(ka)}-(1+h_{\rm eff}^{\rm A})\right \} \nonumber \\
&&\hspace{-1cm}-2zJ_0S\sum_k\left \{[\sqrt{(1+h_{\rm eff}^{\rm A})^2-\epsilon^2\cos^2(ka)}(\alpha_k^{*}\alpha_k+\beta_k^*\beta_k)\right \}, \nonumber \\
\end{eqnarray}
and the self-consistent equations,
\begin{eqnarray}
\label{sceq}
m_\lambda&=&S\nonumber \\
&&\hspace{-1.5cm}-\frac{2}{N}\sum_{k}\langle \cosh^2\theta_k\alpha_k^*\alpha_k+\sinh^2\theta_k\beta_k^*\beta_k +\sinh^2\theta_k\rangle, 
\end{eqnarray}
where
\begin{eqnarray}
\tanh 2\theta_k =\frac{\epsilon\cos (ka)}{1+h_{\rm eff}^{\rm \lambda}},
\end{eqnarray}

\begin{eqnarray}
h_{\rm eff}^{\rm \lambda}=\frac{H_{\rm eff}^{\rm \lambda}}{2zJ_0S},
\end{eqnarray}
and $z=2$ is a number of the nearest-neighbor spins in the $c$-axis.
The first and second terms in the sum on Eq. \ref{sceq} yield the temperature dependence of the magnetization. 
 We ignore these terms because the effect of the temperature is much smaller than exchange interaction $J_0$ $(k_BT/J_0\sim 0.05)$ in the condition of NMR experiment,

The third term in the summation yields the spin reductions of the ground state.
The self-consistent equations of the spin reductions
from the FR ground state are obtained to be,
\begin{eqnarray}
\Delta S_{\rm \lambda}=\frac{1}{2\pi}\int_{-\pi}^{\pi}dk \sinh^2\left [\frac{1}{2}\tanh^{-1}\left (\frac{\epsilon\cos(k)}{1+h_{\rm \lambda}}\right)\right ].
\end{eqnarray}
We solve these equations numerically. We set the second nearest neighbor interaction as $J_2=0$ because $J_2$ has very small influence on the results when its value is sufficiently smaller than $J_1$. The dependence of the spin reductions of each sublattice on $J_1/J_0$ and $\epsilon$ are plotted in Fig.~\ref{fig4}. The amount of spin reductions of the opposite sublattice decreases as $J_1/J_0$ increases. The amounts of spin contractions of the two parallel sublattices also decrease as $J_1/J_0$ increases, however, this variation is smaller than that of the opposite sublattice. 
The molecular fields make the quantum fluctuations be canceled in the parallel sublattice.
As a result, the opposite and parallel sublattice is asymmetric.

\begin{figure}
\begin{center}
\includegraphics[scale=0.5]{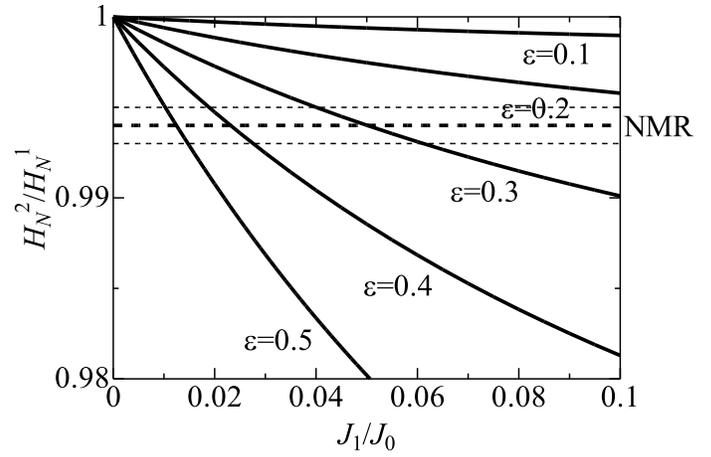}
\caption[]{$J_1/J_0$ and $\epsilon$ dependence of the ratio of the parallel sublattice magnetization and opposite sublattice magnetization. The dashed line represents the ratio of the two hyperfine fields of $^{59}$Co nucleus from the $^{59}$Co NMR results of CsCoBr$_3$.  \label{fig5}}
\end{center}
\end{figure}

The dependence of the ratio of the parallel sublattice and opposite sublattice magnetizations on $J_1/J_0$ and $\epsilon$
are shown in Fig.~\ref{fig5}, in order to compare our results to the NMR results.
The dashed line represents the ratio of the two hyperfine fields of the $^{59}$Co nucleus from the $^{59}$Co NMR results of CsCoBr$_3$. 
Assuming that the two hyperfine fields of $^{59}$Co appear only due to the spin reductions of quantum fluctuation, we obtain the anisotropy parameter $\epsilon$ to be approximately $0.3\sim 0.5$ when $J_1/J_0=0.01\sim 0.06$ in the case of CsCoBr$_3$. This value is slightly larger than the values estimated from the experiments: $\epsilon =0.1\sim 0.2$. \cite{epsilon}
However, even a large hyperfine field of $^{59}$Co of CsCoBr$_3$ is smaller than that of CsCoCl$_3$. The most likely explanation for this is that the quantum fluctuation of the ground state of CsCoBr$_3$ is greater than that of CsCoCl$_3$. Therefore, it is reasonable to suppose that the anisotropy parameter $\epsilon$ of CsCoBr$_3$ is greater than that of CsCoCl$_3$.

Moreover, we observe that the ratio approaches 1 when the $J_1/J_0$ decreases.
The value of $J_1/J_0$ of CsCoCl$_3$ that is experimentally estimated 
is much lower than that of CsCoBr$_3$.
As a result, it can be considered that the $^{59}$Co NMR spectra of CsCoCl$_3$
have sharp peaks which are not observed in CsCoBr$_3$. 

\section{Conclusions}
In summary, we study the spin reductions of the ground state on the stacked triangular antiferromagnet.
We observe the sublattice asymmetry of the spin reductions.
This sublattice asymmetry gives a good qualitative explanation of the NMR results of CsCoBr3; further, there is scope to investigate this qualitatively. Another possibility where a spin may be canted by an interaction is not considered in the present paper. It is necessary to consider the asymmetry of spin reductions for quantitative discussions even if spins are canted.

\section*{Acknowledgment}

The author wishes to thank Dr. Y. Nishiwaki, Dr. H. Watanabe, Prof. S. Miyashita and Prof. S. Maegawa for the fruitful discussions.

\begin{table}
\caption{}{$^{59}$ Co NMR hyperfine fields of CsCoBr$_3$ and CsCoCl$_3$ (Refs. \citen{uyeda} and \citen{tkubo}).\label{table1}}

\begin{tabular}{l|c|c}
\hline
 & CsCoBr$_3$ & CsCoCl$_3$ \\
\hline 
$H_N^1$ [kOe] & $480.0\pm 0.5$ & $499.38\pm 0.05$ \\
$H_N^2$ [kOe] & $477.0\pm 0.5$ & \\
\hline
\end{tabular}
\end{table}

\end{document}